# Base Selection and Transmission Synchronization Algorithm in Quantum Cryptography

Cătălin Anghel

*University "Dunărea de Jos" of Galaţi  
Galaţi, Romania (e-mail: catalin.anghel@ugal.ro).*

**Abstract**: One Achilles heal of classical cryptographic communication systems is that secret communication can only take place after a key is communicated in secret over a totally secure communication channel. Here comes quantum key distribution which takes advantage of certain phenomena that occur at the subatomic level, so that any attempt by an enemy to obtain the bits in a key not only fails, but gets detected as well. This paper proposes the idea of on algorithm, intended to be a quantum network algorithm named Base Selection and Transmission Synchronization - BSTS, which can realize a perfectly secure communication between two computers.

## 1. INTRODUCTION

Cryptography (derived from the Greek words *kryptos* and *graphein* meaning *hidden writing*) is the science of codes and ciphers. A cipher is essentially a cryptographic algorithm which is used to convert a message, known as the plaintext, into unreadable ciphertext. The message can then be safely transmitted without fear of letting sensitive information fall into the hands of the enemy.

## 2. QUANTUM CRYPTOGRAPHY - PRINCIPLES

Cryptographic algorithms, used in our days, are founded on complexity of the mathematical algorithms, but computers become faster and faster and to break an encrypted message becomes a matter of computational power. Consequently, efforts have been made to establish new foundations for cryptography. One of these efforts has lead to the development of quantum cryptography, whose security relies not on assumptions about computer power, but on the laws of quantum physics.

The results in quantum cryptography are based on the *Heisenberg uncertainty principle* of quantum mechanics. Using standard *Dirac notation* [4], this principle can be stated as follows:

For any two quantum mechanical observables *A* and *B* we have :

$$\langle (\Delta A)^2 \rangle \langle (\Delta B)^2 \rangle \geq \frac{1}{4} \| \langle [A,B] \rangle \|^2 \quad (1)$$

where $\Delta A = A - \langle A \rangle$ and $\Delta B = B - \langle B \rangle$

and $[A,B] = AB - BA$

Thus, $\langle (\Delta A)^2 \rangle$ and $\langle (\Delta B)^2 \rangle$ are standard deviations which measure the uncertainty of observables *A* and *B*. For observables A and B, such that $[A,B] \neq 0$, reducing the uncertainty $\langle (\Delta A)^2 \rangle$ of observable A forces the uncertainty $\langle (\Delta B)^2 \rangle$ of observable B to increase, and vice versa. Thus the observables *A* and *B* can not be simultaneously measured to arbitrary precision. Measuring one of the observables interferes with the measurement of the other.

So, using quantum physics phenomena, we can build a perfectly secure key distribution system - this is known as quantum key distribution (QKD). The keys produced using QKD are guaranteed to be secret – as is proved by BB84 protocol [1, 3], and may be used in conjunction with any classical cryptographic system (CCS) [5], such as *one-time pad*.

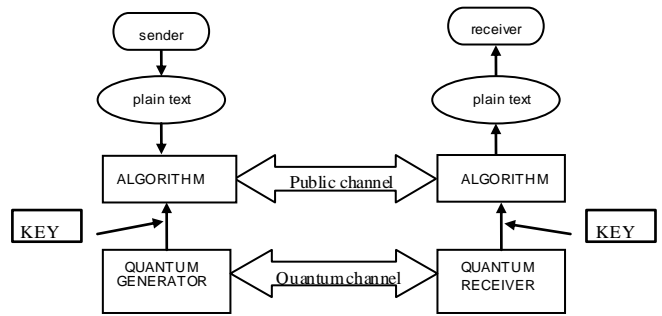

Fig 1. A quantum cryptographic communication system

## 3. QUANTUM KEY DISTRIBUTION

Electromagnetic waves such as light waves can exhibit the phenomenon of polarization, in which the direction of the electric field vibrations is constant or varies in some definite way. A polarization filter is a material that allows only light of a specified polarization direction to pass. Information about the photon's polarization can be determined by using a photon detector to determine whether it passed through a filter. In quantum key distribution, any attempt of an enemy to obtain the bits in a key not only fails, but gets detected as well. Specifically, each bit in a key corresponds to the state of a particular particle, such as the polarization of a photon – named quantum bit (qbit).

The sender of a key has to prepare a sequence of polarized photons - qbits, which are sent to the receiver through an optical fiber. In order to obtain the key represented by a given sequence of photons, the receiver must make a series of measurements using a set of polarization filters.

A photon can be polarized *rectilinear* ($0^o$, $90^o$), *diagonal* ($45^o$, $135^o$) and *circular* (left - spinL, right - *spinR*).

The process of mapping a sequence of bits to a sequence of rectilinearly, diagonally or circularly polarized photons is referred to as conjugate coding, while the rectilinear, diagonal and circular polarization are known as conjugate variables. Quantum theory stipulates that it is impossible to measure the values of any pair of conjugate variables simultaneously. The same impossibility applies to rectilinear, diagonal and circular polarization for photons. For example, if someone tries to measure a rectilinearly polarized photon with respect to the diagonal, all information about the photons rectilinear polarization is lost.

## 4. BB84 ALGORITHM OF QKD

BB84 is the first known quantum key distribution scheme, named after the original paper by Bennett and Brassard, published in 1984. BB84 allows two parties, *Sender* and *Receiver*, to establish a secret, common key sequence using polarized photons - qbits.

*The steps in the procedure are listed below:*

1. Sender generates a random binary sequence s.

2. Sender chooses which type of photon to use (rectilinearly polarized, "R", or diagonally polarized, "D") in order to represent each bit in *s*. Let *b* denote the sequence of each polarization base.

3. Sender uses specialized equipment, including a light source and a set of polarisers, to create a sequence p of polarized photons - qbits whose polarization directions represent the bits in s.

4. Sender sends the qbits p to Receiver over an optical fiber.

5. For each qbit received, Receiver makes a guess of which base is polarized : rectilinearly or diagonally, and sets up his measurement device accordingly. Let b' denote his choices of basis.

6. Receiver measures each qbit with respect to the basis chosen in step 5, producing a new sequence of bits s'.

7. Sender and Receiver communicate over a classical, possibly public channel. Specifically, Sender tells Receiver the choice of basis for each bit, and Receiver tells Sender whether he made the same choice. The bits for which Sender and Receiver have used different bases are discarded from s and s'.

*Detecting Eavesdropper's presence :*

For the $i^{th}$ bit chosen by *Sender*, *s[i]*, there will correspond a choice of polarization basis, *b[i]*, which is used to encode the bit to a photon. If *Receiver* 's chosen measurement basis is *b'[i]* and the outcome of his measurement is *s'[i]*, then:

$$b'[i] = b[i] \text{ should imply } s'[i] = s[i]$$

If an *Eavesdropper* tries to obtain any information about *s[i]*, a disturbance will result and, even if *Receiver* and *Sender* 's bases match, $s'[i] \neq s[i]$. This allows *Sender* and *Receiver* to detect the *Eavesdropper*'s presence, and to reschedule their communications accordingly.

## 5. SECRET KEY RECONCILIATION

The basic BB84 procedure is incomplete because : whether an *Eavesdropper* is present or not, there will still be errors in *Receiver* 's key sequence. The final step of BB84, which was described above merely as a comparison of encoding and measurement bases, is usually much more elaborate. There are two parts involved : *secret key reconciliation* [3] and *privacy amplification* [3, 2].

The process of Secret Key Reconciliation is a special error correction procedure that eliminates :

- errors due to incorrect choices of measurement basis;
- errors induced by *Eavesdropper*;
- errors due to channel noise, if any exist.

Reconciliation is performed as an interactive binary search for errors. *Sender* and *Receiver* divide their bit sequences into blocks and compare each other's parity for each block. Whenever their respective parities for any given block do not match, they divide it into smaller blocks and compare parities again, repeating this process until the exact location of the error is found. When an error has been located, *Sender* and *Receiver* discard the corresponding bit. During this process, *Sender* and *Receiver* can communicate over a classical channel, which is by definition insecure and accessible to an *Eavesdropper*.

## 6. PRIVACY AMPLIFICATION

Since valuable information about the key may be obtained by an *Eavesdropper* during reconciliation, *Sender* and *Receiver* must perform a final step in order to establish a perfectly secret key : this is the process of *privacy amplification*.

The process of reconciliation results in a bit sequence which is common to *Sender* and *Receiver*, but some of its bits may be known by an eavesdropper who has tapped the classical channel. To eliminate this "leaked" information, *Sender* and

*Receiver* must apply, in common, a binary transformation (usually, a random permutation) to their sequences, and discard a subset of bits from the result.

The objective of this step is to minimize the quantity of correct information that the *Eavesdropper* may have obtained about *Sender* and *Receiver* 's common bit sequence.

At the end of the privacy amplification procedure, *Sender* and *Receiver* 's bit sequences may be shown to be identical and absolutely secret, with arbitrarily high probability.

## 7. BASE SELECTION AND TRANSMISSION SYNCHRONIZATION PRINCIPLE

The novelty introduced by this paper is Base Selection and Transmission Synchronization algorithm – named BSTS algorithm.

After the Secret Key Reconciliation process, the bit sequence obtained is too small comparing with the initial binary sequence *s*. As an alternative to *privacy amplification*, we propose a new algorithm, intended to be a quantum network algorithm named Base Selection and Transmission Synchronization - BSTS, which can realize a perfectly secure communication between two computers.

We use the bit sequence obtained after the Secret Key Reconciliation process – named *primary key*, which is common to *Sender* and *Receiver*, to establish a totally new communication protocol.

In order to create the quantum communication process between *Sender* and *Receiver*, we need to use some special devices that can transmit, polarize and receive photons and without a computer with special software this communication can not be done. This software can be implemented in order to transmit, receive and analyse the photon transmission between *Sender* and *Receiver*.

This new communication protocol will establish, accordingly to the *primary key*, which pair of bases, between *rectilinear* ($0°$, $90°$), *diagonal* ($45°$, $135°$) and *circular* (left - spinL, right - spinR), will be used for photons polarizations. Accordingly to the *primary key*, the communication protocol will establish also the polarization base for each photon that has to be transmitted, so the *Sender* and *Receiver* will know for each particular photon the polarization base – this process will be named *base selection*. Finally, accordingly to the *primary key*, the communication protocol will establish the moments in time when *Sender* will send a photon and *Receiver* will read the photon – this process will be named *transmission synchronization*.

In the rest of the time *Sender* will send out fake arbitrary polarized photons which will be ignored by the *Receiver*.

## 8. BASE SELECTION AND TRANSMISSION SYNCHRONIZATION ALGORITHM

Accordingly to the *primary key*, which is common to *Sender* and *Receiver*, we propose the following new algorithm:

1. First two bits from *primary key* will be used to establish the two polarization bases which will be used to polarize the photons for current transmission:

**Table 1. Base selection**

| Bit 1 | Bit 2 | Base 1 | Base 2 |
|---|---|---|---|
| 0 | 0 | **R**(rectilinear) | **D**(diagonal) |
| 0 | 1 | **R**(rectilinear) | **C**(circular) |
| 1 | 0 | **C**(circular) | **D**(diagonal) |
| 1 | 1 | **R**(rectilinear) | **D**(diagonal) |

2. Next four bits from *primary key* will be a number that it's conversion in 10 bases will represent the timing interval expressed in milliseconds. Accordingly to the timing interval the *Sender* will send photons and *Receiver* will read photons. *Sender* can communicate to *Receiver* thru the public channel the moments of start and the end of transmission.

**Table 2. Time**

| Bit 3 | Bit 4 | Bit 5 | Bit 6 | Time (ms) |
|---|---|---|---|---|
| 0 | 0 | 0 | 0 | 1 |
| 0 | 0 | 0 | 1 | 2 |
| …………………………………………………… | | | | |
| 1 | 1 | 1 | 1 | 16 |

3. The remaining bits from *primary key* will represent the polarization base of each photon that we have to transmit :

**Table 3. Base**

| Bit | 0 | 1 |
|---|---|---|
| Polarization | Base 1 | Base 2 |

After the last bit from *primary key* the process of polarization will be continued from the bit number 7 from *primary key*.

## 9. EXAMPLE

The bits in *primary key* are : 01100110011

**Table 4. Bits in *primary key***

| Nr | 1 | 2 | 3 | 4 | 5 | 6 | 7 | 8 | 9 | 10 | 11 | … |
|---|---|---|---|---|---|---|---|---|---|---|---|---|
| Bit | 0 | 1 | 1 | 0 | 0 | 1 | 1 | 0 | 0 | 1 | 1 | … |

*Sender* and *Receiver* have the same bits in *primary key* and they make the same actions:

1. Determine the two bases for photons polarization

**Table 5. Base selection**

| Bit 1 | Bit 2 | Base 1 | Base 2 |
|---|---|---|---|
| 0 | 1 | **R**(rectilinear) | **C**(circular) |

They will use for photons polarization the rectilinear and circular bases.

2. Determine the interval for sending and receiving real photons

**Table 6. Bits 3,4,5,6 from *primary key***

| Bit 3 | Bit 4 | Bit 5 | Bit 6 | Time (ms) |
|---|---|---|---|---|
| 1 | 0 | 0 | 1 | 9 |

*Sender* will transmit the fake photons all the time but from 9 to 9 ms it will transmit the real photons to the *Receiver*. Also the *Receiver* will read photons only from 9 to 9 ms.

3. Determine each photon polarization base

**Table 7. Selection of base**

| Bit | 0 | 1 |
|---|---|---|
| Polarization | Base 1 | Base 2 |

**Table 8. Photons polarizations bases**

| Nr | 1 | 2 | 3 | 4 | 5 | 6 | 7 | 8 | 9 | 10 | 11 | … |
|---|---|---|---|---|---|---|---|---|---|---|---|---|
| Bit | 0 | 1 | 1 | 0 | 0 | 1 | 1 | 0 | 0 | 1 | 1 | … |
| Base | | | | | | | C | R | R | C | C | … |

So, the first bit that we have to transmit will be represented by a photon polarized in circular base, the second one will be represented by a photon polarized in rectilinearly base, and so on. After the last bit from *primary key* the process of polarization will be continued from the bit number 7.

## 10. CONCLUSION

Quantum cryptography and especially Quantum Key Distribution (QKD) enables Secret Key Establishment between two users, using a combination of a classical channel and a quantum channel, such as an optical fibre link. The essential interest of QKD is that any eavesdropping on the line can be detected. This property leads to a cryptographic properties that cannot be obtained by classical techniques and also allows to perform Key Establishment with an extremely high security standard which is known as unconditional security.

In this paper, the intention was to propose the idea of on algorithm that can realize a quantum transmission between two computers. This algorithm, based on the result of the *Secret Key Reconciliation* process of the BB84, can assure a high security communication between two users.

The *base selection and transmission synchronization algorithm*, intend to be a quantum network algorithm that can realize a quantum connection between two computers.

For future directions, the intention is to build a computer program that use the *base selection and transmission synchronization algorithm*.